\documentclass[12pt]{iopart} 

\usepackage[T1]{fontenc}    
\usepackage[utf8]{inputenc} 

\usepackage{cite}
\usepackage{amsfonts}
\usepackage{amssymb}
\usepackage{graphicx}
\usepackage{subfigure}
\usepackage{color}

\usepackage{feynmp}

  \DeclareGraphicsExtensions{.pstex,.eps}

\makeatletter%
\def\fps@figure{tbp}%
\def\fps@table{tbp}%
\makeatother%

\begin{document}

\title[Crossover from directed percolation to mean field behavior]
      {Crossover from directed percolation to mean field behavior in the diffusive contact process}
\author{Andreas Messer and Haye Hinrichsen}
\address{Universit\"at W\"urzburg\\
	 Fakult\"at f\"ur Physik und Astronomie\\
         D-97074 W\"urzburg, Germany}
\ead{hinrichsen@physik.uni-wuerzburg.de}

\begin{abstract}
Recently Dantas, Oliveira and Stilck [J. Stat. Mech. (2007) P08009] studied how the one-dimensional diffusive contact process crosses over from the critical behavior of directed percolation to an effective mean field behaviour when the diffusion rate is sent to infinity. They showed that this crossover can be described in terms of a crossover exponent $\phi$, finding the boundaries $3 \leq \phi \leq 4$ in one spatial dimension. In the present work we refine and extend this result up to four spatial dimensions by a 
field-theoretic calculation and extensive numerical simulations.
\end{abstract}

\pacs{05.50.+q, 05.70.Ln, 64.60.Ht}

\parskip 2mm 

\section{Introduction}

In non-equilibrium statistical mechanics, the study of phase transition from fluctuating phases into absorbing states continues to be a very active field~\cite{MarroDickman99}. One of the most important universality classes of such transitions is directed percolation (DP)~\cite{Kinzel85,Hinrichsen00,Odor04,Lubeck04}, which is well understood and can be described in terms of a renormalizable field theory~\cite{Cardy96,Taeuber07}. Although this universality class plays a paradigmatic role as the Ising class in equilibrium statistical mechanics, it was primarily of theoretical interest since experimental realizations seemed to be very difficult~\cite{Hinrichsen00c}. However, very recently Takeuchi {\it et al.}~\cite{TakeuchiEtAl07} performed an experiment on the basis of turbulent liquid crystals, where the critical exponents of directed percolation could be measured.

Directed percolation is most easily introduced as a spreading process of diffusing particles on a lattice which multiply ($A\to 2A$) and self-annihilate ($A\to \emptyset$). Moreover, the density of particles is effectively limited by an exclusion principle. One of the most important models with random-sequential updates is the contact process~\cite{MarroDickman99} which is controlled by the rate for offspring production $\lambda$ with a critical point $\lambda_c$. The DP universality class is characterized by three critical exponents $(\beta,\nu_\parallel,\nu_\perp)$ which depend only on the dimensionality of the system. In $d=1,2,3$ spatial dimensions the exponents take non-trivial, probably irrational values. In $d>4$ dimensions, however, the mean-field exponents $\beta=\nu_\parallel=1$ and $\nu_\perp=1/2$ become exact. Right at the upper critical dimension $d_c=4$ there are additional logarithmic corrections.

The crossover to mean field behavior for $d>4$ is related to the fact that the diffuse mixing becomes more efficient in high dimensions, suppressing fluctuation effects. An alternative way to enhance diffusive mixing, on which we will focus in the present work, is to increase the diffusion constant $D$ in a low-dimensional system. In most models such as the contact process, diffusion is effectively included by the circumstance that a particle creates offspring at an empty randomly chosen nearest-neighbour site. In such models the effective diffusion rate is fixed but it is straight forward to add explicit diffusion of solitary particles so that the diffusion rate can be controlled.

By varying the diffusion constant in the contact process, one expects the following phenomenology. Clearly, in the limit $D \to \infty$, where all sites are mutually coupled, mean field theory becomes exact. For a large but finite diffusion constant, however, one expects a crossover at a typical length scale $\xi^{(c)}_\perp$ and an associated typical time scale $\xi^{(c)}_\parallel$ which grow with $D$. Below these scales, the process exhibts an effective mean-field behavior while it crosses over to DP on scales larger than $\xi^{(c)}_\perp$ and $\xi^{(c)}_\parallel$. In addition, the critical rate for offspring production, $\lambda_c(D)$, decreases with $D$ and reaches the analytically known mean field value $\lambda_c(\infty)$ in the limit $D\to \infty$.

The influence of the diffusion rate on $\lambda_c$ was first studied by Konno~\cite{Kon95} from a mathematical point of view. He pointed out that the critical threshold $\lambda_c(D)$ approaches the asymptotic value $\lambda_c(\infty)$ 
algebraically as
\begin{equation}
\lambda_c(D)-\lambda_c(\infty) \;\sim\; D^{-1/\phi}\,,
\end{equation}
where the crossover exponent $\phi=3$ in one, $\phi=1 +\ log.\ corrections$ in two and $\phi=1$ in more than two spatial dimensions. More recently, by series
expansion and partial differential approximants, Dantas {\it et al.}~\cite{DantasEtAl07} found the boundaries $3 \leq \phi \leq 4$ for this exponent. In the present work we describe a pure
field theoretic approach which allows an easy calculation of the crossover exponents. 
Furthermore we present numerical results for one to four spatial dimensions.

\section{Field-theoretical approach}
\label{sec:analytical}
\fmffile{fmfdefault}

\def\part#1{\frac{\partial}{\partial #1}}
\def\mean#1{\left< #1 \right>}
\def\order#1{\mathcal{O}\left(#1\right)}
\def\funct#1#2{ #1 \left[ #2 \right]}
\def\functd#1{\frac{\delta}{\delta #1}}

\def\text#1{{\rm #1}}

\def\tp{\tilde{\phi}}
\def\p{\phi\vphantom\tp}

\def\tj{\tilde{J}}
\def\j{J\vphantom\tj}

\def\eqfmfgraph#1#2#3#4{%
\newdimen\w \w=#3pt \multiply\w by 2 \advance\w by #1pt %
\newdimen\h \h=#3pt \multiply\h by 2 \advance\h by #2pt %
\parbox{\w}{\rule[-#3pt]{0pt}{\h}\centering\begin{fmfgraph}(#1,#2)%
#4\end{fmfgraph}}}
\def\eqfmfgraphstar#1#2#3#4{%
\newdimen\w \w=#3pt \multiply\w by 2 \advance\w by #1pt %
\newdimen\h \h=#3pt \multiply\h by 2 \advance\h by #2pt %
\parbox{\w}{\rule[-#3pt]{0pt}{\h}\centering\begin{fmfgraph*}(#1,#2)%
#4\end{fmfgraph*}}}

The diffusive DP model is most easily described by a contact process defined 
by the three micro processes (i) creation
of a particle on, (ii) hopping to one of the neighboring places, and 
(iii) death of a particle which take place with rates $\lambda$, $D$ and $1$ 
respectively. These dynamic laws result in the coarse grained Langevin equation
\cite{Janssen81}
\begin{equation}\label{eq:langevin}
\part t \rho\left(\vec{x},t \right) = \frac{D}{2 d} \Delta \rho \left(\vec{x},t \right) 
+ \left( \lambda - 1 \right) \rho \left(\vec{x},t \right) - \lambda \rho^2\left(\vec{x},t \right)
+ \xi \left(\vec{x},t \right)
\end{equation}
with spatial dimension $d$, density of particles 
$\rho\left(\vec{x},t \right)$ and multiplicative noise 
$\xi \left(\vec{x},t \right)$ with the correlations
\begin{equation}
\mean{ \xi\left(\vec{x},t \right)\xi\left(\vec{y}, t' \right)}%
\sim \rho\left(\vec{x},t \right) \delta^d\left(\vec{x} - \vec{y} \right)\delta\left(t-t'\right) {\rm .}
\end{equation}
Based on this equation one can set up a field theory by eliminating the 
noise and introducing a response field. After appropriate 
rescaling of fields and parameters, the field theoretic action reads
(see \cite{Taeuber07} and references therein)
\begin{equation}\label{eq:action}
\mathcal{S} = \int d^dx dt \frac{1}{2} \left( \tp \part t \p - \p \part t \tp \right) +  \frac{D}{2 d} \left(\nabla \tp\right) \left(\nabla \p \right) - \tp \kappa \p + g \left( \tp \phi^2 - \tp^2 \p \right)
\end{equation}
with density (response) fields $\p$ ($\tp$), control parameter 
$\kappa = \lambda-1$ and coupling $g$. As seen from \eref{eq:action} 
the field theory of DP involves a directed propagator 
and two cubic interaction vertexes with the same coupling strength $g$. Using
these Feynman rules, one can calculate the next order (one loop) corrections
to propagator and vertexes. For instance, the inverse propagator to 
one loop order is given by
\begin{equation}
\Gamma^{\left(1,1\right)}\left(k, \omega \right) =
\eqfmfgraph{40}{20}{5}{%
\fmfleft{i}
\fmfright{o}
\fmf{dashes_arrow}{i,o}
} %
- \eqfmfgraph{40}{20}{5}{%
\fmfleft{i}
\fmfright{o}
\fmf{dashes_arrow}{i,v1}
\fmf{dashes_arrow,right,tension=1/4}{v1,v2}
\fmf{dashes_arrow,left,tension=1/4}{v1,v2}
\fmf{dashes_arrow}{v2,o}
\fmfdot{v1,v2}}
+\order{g^4}\text{.}
\end{equation}
Naturally these loop corrections lead to divergent contributions, which
receive a physical meaning upon applying a regularization and renormalization
procedure. The divergences then appear as corrections to the meanfield scaling
exponents
\begin{equation}
\nu_\perp = \frac{1}{2} + \frac{\epsilon}{16}+ \order \epsilon^2
,\quad \nu_\parallel = 1 + \frac{\epsilon}{12}+ \order \epsilon^2
,\quad \beta = 1 - \frac{\epsilon}{6}+ \order \epsilon^2
\end{equation}
where $\epsilon=4-d$. As expected, the critical exponents do not depend
on the value of the diffusion rate. 

To get an insight how the diffusion rate $D$ changes the critical creation rate $\lambda_c$, let
us consider the field theoretic calculation at the beginning. Due to convergence problems
upon applying dimensional regularization, one is forced to replace the parameter $\kappa$ by 
mass $m^2=\kappa-\kappa_c\left(D\right)$ where
\begin{equation}\label{eq:kappac}
\kappa_c = g^2 \int \frac{d^d q}{\frac{D}{2 d} q^2 - \kappa_c}%
\quad \text{with\ } \sqrt{2 d \kappa_c / D} < \left | q \right| < \Omega
\end{equation}
is chosen in a way that $m$ becomes zero at criticallity \cite{Taeuber07}.
The lower boundary is given by the fact, that for $q$  smaller than
$\sqrt{2 d \kappa_c / D}$ the propagator becomes zero. The upper boundary $\Omega$
is a cutoff scale in momentum space and should be sent to infinity under continued
renormalization.
Inserting the $d$-dimensional surface element and expanding the right hand side 
of \eref{eq:kappac} in powers of $q$ one obtains an integral over a geometric series
\begin{equation}\label{eq:kcexpansion}
\kappa_c = \frac{4 d \pi^{d/2} g^2}{D \Gamma\left(d/2\right)}%
\int^\Omega_{\sqrt{2 d \kappa_c / D}} d q\,  q^{d-3} \sum_{i=0}^\infty \left(\frac{2 d q^{-2}\kappa_c}{D}\right)^i{\rm .}
\end{equation}
By dimensional analysis one can show that for $d < 2$ only the lower boundary 
and for $d > 2$ only the upper boundary contributes to the integral. As we are 
interested in the asymptotic behavior for large $D$, we may approximate the sum
by the leading term $1$.
Integrating the remaining part then yields the crossover behavior for 
different spatial dimension
\begin{equation}\label{eq:lambdac}
\lambda_c\left(D\right)-1 \sim \left\{%
\begin{array}{cl}
D^{- \frac{d}{4-d}} & \text{for\ } d < 2 \\
\frac{\log D}{D} & \text{for\ } d = 2\\
D^{-1} & \text{for\ } d > 2
\end{array}%
\right.
\end{equation}
These results compare very well to the predictions made 
in \cite{Kon95} but here they are obtained in a much simpler and direct way. 
From \eref{eq:lambdac} we read off the crossover exponents
\begin{equation}
\phi\left(d\right) = \left\{%
\begin{array}{cl}
3 &  \text{for\ } d = 1 \\
1 + log.\ corrections  & \text{for\ } d = 2 \\
1 &  \text{for\ } d \geq 3  
\end{array}%
\right.
\end{equation}

\section{Numerical results}
\begin{figure}
\centering
\includegraphics{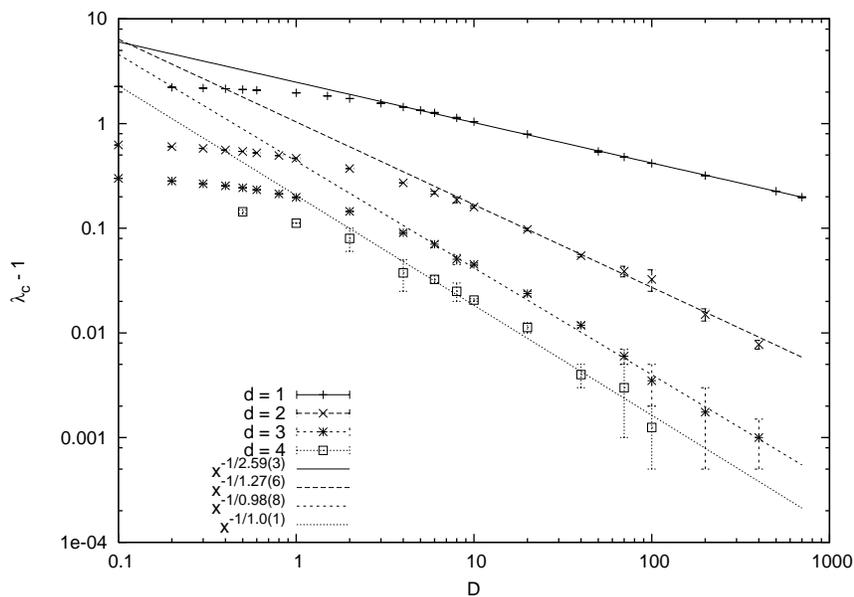}
\caption{Log-Log plot of $\lambda_c\left(D\right)$ derived by Monte-Carlo simulation. 
The straight lines show power laws, fitted to the large $D$ part of the numerical data.}
\label{fig:log}
\end{figure}

In order to substantiate these analytical results, we simulated
the diffusive contact process on periodic $d$-dimensional lattices
using the following dynamic rules: Select a random lattice site and if
there is a particle perform one of the following moves:
\begin{enumerate}
\item remove the selected particle with probability $1/\left(1+D+\lambda\right)$,
\item create a particle on a randomly chosen neighbouring site with prob. 
$\lambda/\left(1+D+\lambda\right)$ if this site is empty, or
\item move the particle to randomly selected neighbouring site with prob. 
$D/\left(1+D+\lambda\right)$ if this site is empty.
\end{enumerate}
For various values of the diffusion rate $D$ we measured the particle density in a system
starting with a fully occupied lattice and averaged over at least 20 independent runs.
By varying the creation rate $\lambda$ and searching for a power law decay we determined
the critical threshold $\lambda_c$ as a function of $D$. These critical lines
for one, two, three and four spatial dimensions are shown in figure \ref{fig:log}. 
The plots clearly indicate the expected 
power law behavior for large diffusion rates and the exponents are in
good agreement with our analytical results. For small $D$, however, the
observed deviations from a pure power law are not very surprising 
because the diffusion length becomes comparable to the lattice spacing.
\begin{figure}
\centering
\includegraphics{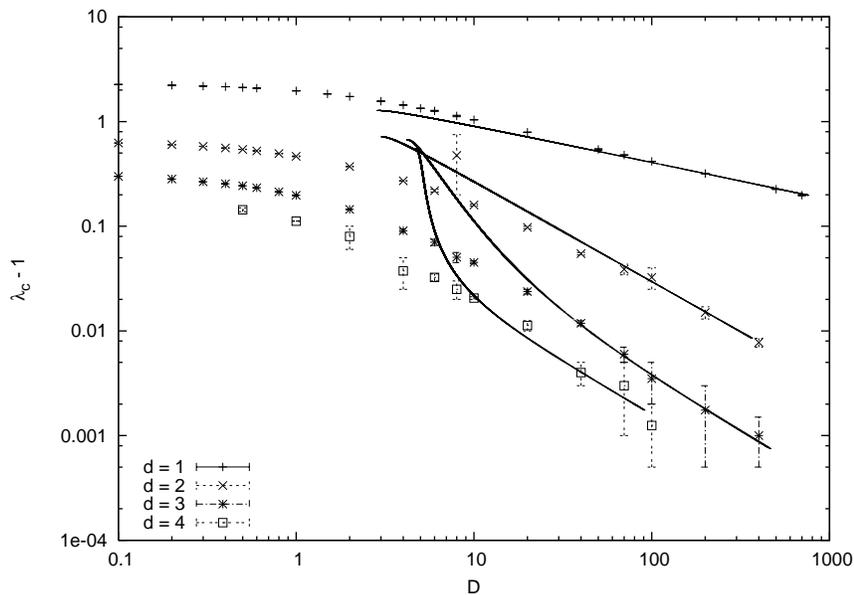}
\caption{The same simulation results as in figure \ref{fig:log} compared to 
the numerically integrated predictions of eq. \eref{eq:kappac} (solid lines).  
Since the iteration becomes unstable for small diffusion rate, this 
region is not accessible by this approach. }
\label{fig:ndata}
\end{figure}

So far we verified the predicted asymptotic behavior \eref{eq:lambdac}.
In order to investigate the next-leading corrections of the geometric series, 
we integrate \eref{eq:kcexpansion} numerically and compare it with the results
of the simulations (figure  \ref{fig:ndata}). 
Although for small $D$ this solution is plagued by numerical
instabilities rendering the results unusable, the reader should 
notice that for spatial dimension $d=1$ the numerical integration becomes 
almost exact over at least two decades and does not show the lift off behavior
like in other dimensions.
\section{Concluding remarks}
In this paper we have presented a simple field-theoretic calculation and 
extensive numerical simulations in order to investigate how the critical
parameter depend on the diffusion rate of a DP process in various dimensions. Our field-theoretical
result confirms previous results by Konno \cite{Kon95}, derived here
in a much simpler way. Moreover we have presented numerical results
which are in agreement with these predictions, refining previous
results by Dantas {\it et al.} \cite{DantasEtAl07} and extending them to higher
space dimensions up to $d=4$. Especially in one spatial dimension, 
the field-theoretic prediction to one loop order, when evaluated numerically, 
coincides almost perfectly with the Monte Carlo estimates. This is surprising 
since one expects loop corrections to be more relevant in lower dimensions.

\vspace{5mm}
\noindent{\bf References}
\bibliographystyle{unsrt}

\end{document}